# Linguistic Relativity and Programming Languages

Jiahao Chen *

**Abstract**

The use of programming languages can wax and wane across the decades. We examine the split-apply-combine pattern that is common in statistical computing, and consider how its invocation or implementation in languages like MATLAB and APL differ from R/dplyr. The differences in spelling illustrate how the concept of linguistic relativity applies to programming languages in ways that are analogous to human languages. Finally, we discuss how Julia, by being a high performance yet general purpose dynamic language, allows its users to express different abstractions to suit individual preferences.

## 1. Each decade has its own programming languages

Each discipline has its own favorite languages: applied mathematics has MATLAB, web applications have JavaScript, and high-performance computing still uses Fortran, accompanied by Python (Jones et al., 2001) and even Tcl (Phillips et al., 2014). Statistical computing, of course, is associated strongly with R. Yet the boundaries of these seemingly absolute fiefdoms in the kingdom of computing turn out to be surprisingly malleable on the time scale of decades. Statistical computing came to prominence with PL/I in the 70s, APL in the 80s, and XLISP-STAT in the 90s before the relatively modern advent of S and iR (de Leeuw, 2005). In truth, programming languages come and go beyond the time scale of single years or even PhD studentships.

So what makes a programming language suitable for the demands of a scientific discipline like statistical computing? Discussing pros and cons of different languages can get bogged down in absolutist statements about what "can" and "cannot" be done in a language. On some level such statements are absurd, given that all sufficiently complicated programming languages are Turing complete, and therefore have the same computational power in the sense of Turing equivalence. Thus anything that can be done in one Turing-complete language ***must*** be doable in another Turing-complete language. Instead, the answer to the question of suitability must necessarily be ensconced in "softer" concepts about ease of use and expressiveness, concepts that are hard to define precisely but are nonetheless responsible for shaping the adoption of programming languages.

This paper explores how the suitability of programming languages is related to expressiveness: what abstractions exist in a given programming language that map onto ideas that a programmer would want to implement? Closely related is the notion of idiomaticness: would an experienced programmer in a particular language recognize and accept a given piece of code as "idiomatic"?

---

*Computer Science and Artificial Intelligence Laboratory, Massachusetts Institute of Technology, Cambridge, Massachusetts, 02139. Current address: Capital One Financial, 11 West 19th Street, New York, New York 10011. (`jiahao.chen@capitalone.com`)

## 1.1 Linguistic relativity and programming languages

To discuss the suitability of programming languages, I will borrow notions from the study of **human** languages, linguistics. In particular, I will adopt the controversial concept of linguistic relativity (Gumperz and Levinson, 1996), or the Sapir–Whorf hypothesis (Brown, 1976), and argue that the idea that (human) languages determine or even constrain cognition has its relevance to computer languages and programming.

Some of the original writings of Sapir and Whorf seem particularly relevant to the discussion. Here are some choice quotations:

> Human beings do not live in the objective world alone,[...] but are very much at the mercy of the particular language which has become the medium of expression for their society [...] [T]he 'real world' is to a large extent unconsciously built up on the language habits of the group. (Sapir, 1929)

Sapir wrote about human languages, in particular contrasting native American languages like Hopi against Occidental languages like English. Nevertheless, programmers would immediately recognize these words about "language habits" as echoing a parallel worldview in computer languages, being themselves human constructs designed to abstract away unwanted, low level machine details. Some examples of these programming language habits may include:

"Always put data in a data frame"

"Never write for loops; vectorize your code to gain performance."

Some "language habits" can be as banal as choosing between 0- or 1-based indexing, leading to notoriously unproductive flame wars. Others are more subtle. For example, MATLAB treats vectors like column matrices, and therefore allow operations on one-dimensional objects like `v[(2,1)]` indexing with two numbers, which are disallowed in many other languages like C. We can trace the different indexing behaviors back to different mathematical starting points—whereas most languages treat vectors as one-dimensional arrays, i.e. "flat" sequences of numbers, MATLAB by virtue of treating most objects like matrices, assumes that vectors by default are **column** vectors, and therefore behave just matrices with one column.

The fact that such discussions of language habits tend to be emotionally charged is also not a coincidence. Whorf wrote:

> [E]very person [...] carries through life certain naïve but deeply rooted ideas about talking and its relation to thinking. Because of their firm connection with speech habits that have become unconscious and automatic, these notions tend to be intolerant of opposition. (Whorf, 1956b)

Replace "talking" with "writing code", and the analogy between speech and computer programs could hardly be plainer. An experienced programmer, practically by definition, internalizes the boilerplate and design patterns in code as "unconscious and automatic" idioms to be regurgitated on demand (preferably with an editor or IDE that helps reinforce these idioms automatically). To the fluent Java programmer, wrapping everything in a class must be second nature, just as the R user is accustomed to seeing data in a data frame, or whitespace sensitivity to the Pythonista. Allowing for code as a generalization of speech, one could argue that Whorf's observation predicted the very phenomenon of flame wars over programming language design!

In the rest of this paper, I will make use of another insight from Whorf:

Through [linguistic knowledge], the world as seen from the diverse viewpoints of other social groups, that we have thought of as alien, becomes intelligible in new terms. Alienness turns into a new and often clarifying way of looking at things. (Whorf, 1956a)

I will present a simple, stereotypical data science task and show how solutions may be implemented in different computer languages, not all of which may necessarily be familiar to the reader or the typical user of statistical computing. This side-by-side comparison of the familiar and unfamiliar will hopefully aid to highlight similarities and differences between the abstractions expressed within the codes.

## 2. A simple data science task: split, apply, combine

Here is a simple data analysis task that is perhaps emblematic of our house sharing, carpooling times. Suppose I am a data scientist working at a ride sharing company and here are the user ratings data for the last ten trips taken by a particular driver:

| userid | 381 | 1291 | 3992 | 193942 | 9493 | 381 | 3992 | 381 | 3992 | 193942 |
|--------|-----|------|------|--------|------|-----|------|-----|------|---------|
| rating | 5 | 4 | 4 | 4 | 5 | 5 | 5 | 3 | 5 | 4 |

Suppose also that I am interested in working out the average rating given by each unique user. A seasoned R user might immediately recognize this task as a "split-apply-combine" problem, which could be solved using the `dplyr` package as follows (Wickham, 2011):

```
library(dplyr);
userid = c(381, 1291, 3992, 193942, 9493, 381,
  3992, 381, 3992, 193942)
rating = c(5, 4, 4, 4, 5, 5, 5, 3, 5, 4)
mycar = data.frame(rating, userid)
summarize(group_by(mycar, userid), avgrating=mean(rating))
```

```
# A tibble: 5 x 2
   userid avgrating
    <dbl>     <dbl>
1     381  4.333333
2    1291  4.000000
3    3992  4.666667
4    9493  5.000000
5  193942  4.000000
```

The key computation is expressed by the `summarize` function, a higher-order function which combines the data in the `mycar` data frame after being split by (grouped by) `userid` and had the function `mean` applied to each group. This function is provided by `dplyr` and is perfectly well-suited to the split-apply-combine task. Now let's look at how the same idiom can be expressed in other programming languages.

## 3. MATLAB: split-apply-combine on matrices

MATLAB is more often thought of as a language for scientific computing rather than statistical computing. Nevertheless, MATLAB provides a higher order function, `accumarray`, which is perfectly suited to split-apply-combine computations. Even MATLAB seems to admit that `accumarray` is "under-appreciated" (Shure, 2008); nevertheless, the function does the job admirably:

```
userids = [381; 1291; 3992; 193942; 9493; 381; 3992; 381; 3992;
    193942];
ratings = [5; 4; 4; 4; 5; 5; 5; 3; 5; 4];
accumarray(userids,ratings,[],@mean,[],true)

ans =

                (381,1)                           4.3333
               (1291,1)                           4.0000
               (3992,1)                           4.6667
               (9493,1)                           5.0000
             (193942,1)                           4.0000
```

Unlike R, which was designed from the beginning around data frames as one of the fundamental data structures, MATLAB was originally designed with matrices as the sole data structure (Moler, 1980, 1982). While MATLAB as of R2013b now provides tables as a data structure (Shure and Zaranek, 2013), most of the base language is still built around matrices and does not work with tables. `accumarray` is one example of the base language that is designed around matrices.

It can be difficult to determine from the documentation of `accumarray` (The Math-Works, Inc., 2016) that it is a function that solves the split-apply-combine problem. The first complete sentence in the documentation states that `A=accumarray(val,subs)` "returns array A by accumulating elements of vector `val` using the subscripts `subs`". In other words, the most basic use of `accumarray` is to split data in `val` into subsets grouped by the values of their corresponding entries in `val`, with the summation (accumulation) function applied to each subset. It is clear that the function is generalizable: a different function can be specified in the fourth positional argument (MATLAB does not support keyword arguments), but in order to do so, a default argument must be specified for the third argument. Finally, the last argument `true` specifies that the output should be returned as a sparse matrix, otherwise the result would be a $193942 \times 1$ dense matrix with most of the entries zero.

## 4. APL: split-apply-combine using array operations

APL is a language that is built around a single data structure, the array.[1] APL does not provide a standard idiom for solving the split-apply-combine problem; instead, the APL programmer must express split-apply-combine on their own using lower-level array computations.

One implementation of split-apply-combine might look like this:

```
mean←{+/ω÷ρω}
uniqfy←{ω[⍋ω]}∪
∇a←v splitby k
a←{(ω∊k)/v}¨uniqfy k
∇
summarizeby←{(uniqfy ω),[1.5]mean¨(α splitby ω)}
```

---

[1]Some modern implementations such as Dyalog APL (Dyalog, Ltd.,2016) have nonstandard extensions that provide support for object-oriented programming in the form of classes, but we won't consider them here.

```
7   u←381 1291 3992 193942 9493 381 3992 381 3992 193942
8   r←5 4 4 4 5 5 3 5 4
9   ]display r summarizeby u
```

```
┌→─────────────────┐
↓   381 4.333333333│
│  1291 4          │
│  3992 4.666666667│
│  9493 5          │
│193942 4          │
└──────────────────┘
```

APL functions can take at most two arguments; further arguments must be specified in "concealed arguments", which are variables in global scope. This solution chose to hard code the function being applied (`mean`) into the higher-level `summarizeby` function. Furthermore, dyadic APL functions must be specified in place, which suggests the name `summarizeby` rather than `summarize` as being more natural to an English speaker. Notably, the expression `r summarizeby u` aligns with the SVO (subject-verb-object) order of English, which can explain the intuition for why the name `summarizeby` may be preferred.

The primary data structure used to implement the "split" stage is the nested array, which allows for the representation of ragged matrices, or list of lists with unequal lengths. The nested array has long been used to implement statistical libraries in APL (Anscombe, 1981; Friendly and Fox, 1994). In this case, the nested array is generated by the (_ω≡k_)/v¨ expression, which extracts each subset of **v** by common values **k**. The `splitby` function produces the result

```
1   ]display r splitby u
```

```
┌→────────────────────────────────┐
│ ┌→────┐ ┌→┐ ┌→────┐ ┌→┐ ┌→──┐   │
│ │5 5 3│ │4│ │4 5 5│ │5│ │4 4│   │
│ └~────┘ └~┘ └~────┘ └~┘ └~──┘   │
└∊────────────────────────────────┘
```

corresponding to the keys

```
2   ]display uniqfy u
```

```
┌→─────────────────────────────┐
│381 1291 3992 9493 193942      │
└~─────────────────────────────┘
```

The "apply" stage of the computation uses APL's built-in diaeresis (¨, "over each") function. Finally, the "combine" stage produces a two-column matrix, with the first column containing the unique keys and the second column containing the required means. This matrix is produced by the `,[1.5]` expression, which "laminates" its two arguments together.

Notably, the APL solution might be considered by some expert programmers to be not idiomatic. Instead, there is a long tradition of writing concise one-line APL solutions to various problems, and for this problem, such a solution might look like

 `{ω,[1.5]{mean¨{(ω≤u)/r}¨`ω}`{ω[⍋ω]}∪u`

To summarize, the three solutions we have seen so far each have different code organizations and work on different data structures. In tabular form:

|  | R/dplyr | MATLAB | APL |
|---|---|---|---|
| split | summarize | accumarray | splitby |
| apply | (column constructor) | (positional argument) | ¨ |
| combine | summarize | accumarray | summarizeby |
| data structure | data frame (tibble) | matrix | array |

## 5. Julia: flexibility of idioms

The last language I will consider here is Julia, a general-purpose programming language that was originally designed for technical computing (Bezanson, 2015)[2]. Julia is a language that is easy for both compilers and programmers to understand, in the sense that it is a high level dynamic language with language constructs that facilitate compiler analysis and generation of efficient code.

The result of Julia's careful design for expressiveness and performance allows for the different solutions above for the split-apply-combine problem to all be expressible in the same language. The solution to split-apply-combine in R/dplyr derives its performance from an underlying implementation of dplyr in C++ (via Rcpp). In contrast, all the Julia solutions in this section are written in pure Julia. The result is that Julia users can also be Julia developers without necessarily having to learn another language to work on the underlying implementation which does all the heavy lifting. Another advantage is that Julia users can experiment with different idioms and data structures without being confined to just those that have well-optimized implementations, and enjoy reasonable performance in many cases.

For example, the `DataFrames.jl` package provides two different spellings for the R/dplyr approach:

```
using DataFrames
df = DataFrame(
  userids=[381, 1291, 3992, 193942, 9494, 381, 3992, 381, 3992,
  193942], ratings=[5, 4, 4, 4, 5, 5, 5, 3, 5, 4]);
by(df, :userids, df->mean(df[:ratings])) #Same as next line
aggregate(df, :userids, mean)
```

```
5×2 DataFrames.DataFrame
│ Row │ userids │ ratings_mean │
├─────┼─────────┼──────────────┤
│ 1   │ 381     │ 4.33333      │
│ 2   │ 1291    │ 4.0          │
│ 3   │ 3992    │ 4.66667      │
│ 4   │ 9494    │ 5.0          │
│ 5   │ 193942  │ 4.0          │
```

The solution to split-apply-combine in R/dplyr derives its performance from an underlying implementation of dplyr in C++ (via Rcpp). In contrast, the Julia solutions are written in pure Julia. The result is that Julia users can also be Julia developers without necessarily having to learn another language to work on the underlying implementation which does all the heavy lifting. Another advantage is that Julia users can experiment with different idioms and data structures without being confined to just those that have well-optimized implementations, and enjoy reasonable performance in many cases.

---

[2] All the Julia code presented here is written for version 0.5.0.

For example, a user who is unfamiliar with data frames and may prefer instead an array-based solution may choose to implement an APL-style solution using a simple list comprehension:

```
userids=[381, 1291, 3992, 193942, 9494, 381, 3992, 381, 3992,
    193942]
ratings=[5, 4, 4, 4, 5, 5, 5, 3, 5, 4]
[(u, mean(ratings[userids.==u])) for u in unique(userids)]

5-element Array{Tuple{Any,Any},1}:
 (381,4.333333333333333)
 (1291,4.0)
 (3992,4.666666666666667)
 (193942,4.0)
 (9494,5.0)
```

The list comprehension approach is simple and concise, if not particularly fast. Nevertheless, the ability to rewrite a solution multiple different way in Julia allows for Julia users to experiment with different approaches in order to find one with the most acceptable performance. For example, a Julia user may choose to rewrite the list comprehension solution by writing out for loops explicitly, something that is discouraged in all the other languages discussed here (which promote the "vectorize your code" habit). One such implementation example might look like the following:

```
counts = []
uuserids=[]
uratings=[]
for i in eachindex(userids)
    j = findfirst(uuserids, userids[i])
    if j==0 #not found
        push!(uuserids, userids[i])
        push!(uratings, ratings[i])
        push!(counts, 1)
    else #already seen
        uratings[j] += ratings[i]
        counts[j] += 1
    end
end
[uuserids uratings./counts]

5x2 Array{Any,2}:
      381  4.33333
     1291  4.0
     3992  4.66667
   193942  4.0
     9494  5.0
```

Finally, a Julia user may later recognize this code as implementing the split-apply-combine idiom and choose to refactor this code into a reusable function akin to `accumarray`. In fact, `accumarray` itself can be implemented in Julia with just a few lines of code:

```julia
26  function accumarray{Tk,Tv}(
27          subs::AbstractArray{Tk}, val::AbstractArray{Tv},
28          func=sum, fillval=zero(Tv)
29        ; sz=maximum(subs,1), issparse=false)
30
31      counts = Dict{Vector{Tk},Vector{Tv}}()
32      for i = 1:size(subs, 1)
33          counts[subs[i, :]] = #split
34              Tv[get(counts,subs[i, :], Tv[]); val[i...]]
35      end
36
37      A = issparse ? spzeros(sz...) : fill(fillval, sz...) #combine
38      for (key, val) in counts
39          A[key...]= func(val) #apply
40      end
41      return A
42  end
```

Thus even though `accumarray` does not exist in the base Julia library, an experienced user can, without much difficulty, refactor a previous solution into a part that implements the general purpose `accumarray` function. Having multiple options gives Julia users the possibility of choosing the best trade-off between development time and actual execution time when deployed on large data sets.

## 6. Conclusion

Many programming languages dictate one preferred way of doing things, which often encompasses an unwritten set of "language habits" as well as a choice of preferred data structures. In particular, "vectorize your code" is an idiom that is common to R, MATLAB and APL, three languages that are currently or in the past been popular for statistical computing. One reason high level languages have a preferred way of expression is because those idioms over a few preferred data structures have highly optimized implementations and are well integrated into a comprehensive standard library. As a result, these idioms shape the way experienced programmers think about solving problems in those languages.

The promise of Julia, as a high level dynamic language with reasonable performance, allows for multiple approaches to the same problem. An experienced user can, without much difficulty, iterate rapidly from a naïve slow solution to a fast, specialized solution, and further to a fast, general purpose solution.

## Acknowledgments

I would like to thank the Julia developer community for many useful discussions, in particular Alan Edelman for the discussion of how `accumarray` may be implemented in Julia.

## References

Anscombe, F. J. (1981). *Computing in Statistical Science through APL*. Springer New York, New York, NY.

Bezanson, J. W. (2015). *Abstraction in Technical Computing*. PhD thesis, Massachusetts Institute of Technology.

Brown, R. (1976). Reference in memorial tribute to Eric Lenneberg. *Cognition*, 4(2):125 – 153.

de Leeuw, J. (2005). On abandoning XLISP-STAT. *Journal of Statistical Software*, 13(1):1–5.

Dyalog, Ltd. (2016). *Dyalog APL: Language Reference Guide, Dyalog version 15.0*. Bramley, UK.

Friendly, M. and Fox, J. (1994). Using APL2 to create an object-oriented environment for statistical computation. *Journal of Computational and Graphical Statistics*, 3(4):387–407.

Gumperz, J. J. and Levinson, S. C. (1996). *Rethinking Linguistic Relativity*. Studies in the Social and Cultural Foundations of Language 17. Cambridge University Press.

Jones, E., Oliphant, T., Peterson, P., et al. (2001). SciPy: Open source scientific tools for Python.

Moler, C. B. (1980). *MATLAB–an interactive matrix laboratory*. Number 369. Albuquerque, NM.

Moler, C. B. (1982). *Demonstration of a matrix laboratory*, volume 909 of *Lecture Notes in Mathematics*, chapter 8, pages 84–98. Springer, Berlin, Heidelberg.

Phillips, J. C., Stone, J. E., Vandivort, K. L., Armstrong, T. G., Wozniak, J. M., Wilde, M., and Schulten, K. (2014). Petascale Tcl with NAMD, VMD, and Swift/T. In *High Performance Technical Computing in Dynamic Languages (HPTCDL), 2014 First Workshop for*, pages 6–17.

Sapir, E. (1929). The status of linguistics as a science. *Language*, 5(4):207–214.

Shure, L. (2008). Under-appreciated accumarray.

Shure, L. and Zaranek, S. W. (2013). Introduction to the new MATLAB data types in R2013b.

The MathWorks, Inc. (2016). accumarray: construct array with accumulation.

Whorf, B. L. (1956a). Language, mind and reality. In Carroll, J. B., editor, *Language, Thought and Reality: Selected Writings of Benjamin Lee Whorf*. MIT Press & John Wiley, New York.

Whorf, B. L. (1956b). Science and linguistics. In Carroll, J. B., editor, *Language, Thought and Reality: Selected Writings of Benjamin Lee Whorf*. MIT Press & John Wiley, New York.

Wickham, H. (2011). The split-apply-combine strategy for data. *Journal of Statistical Software*, 40(1):1–29.